\documentstyle[12pt]{article}
\begin{document}
\baselineskip 3.9ex

\def\be{\begin{equation}}
\def\ee{\end{equation}}
\def\ba{\begin{array}{l}}
\def\ea{\end{array}}
\def\bea{\begin{eqnarray}}
\def\eea{\end{eqnarray}}
\def\no#1{{\tt   hep-th#1}}
\def\eq#1{(\ref{#1})}
\def\pgap{\vspace{1.5ex}}
\def\sgap{\vspace{3ex}}
\def\ggap{\vspace{10ex}}
\def\gap{\vspace{3ex}}
\def\del{\partial}
\def\om{\omega}
\def\o{{\cal O}}
\def\z{{\vec z}}
\def\re#1{{\bf #1}}
\def\av#1{{\langle  #1 \rangle}}
\def\tM{{\widetilde{\cal M}}}
\def\S{{\cal S}}

\renewcommand\arraystretch{1.5}

\begin{flushright}
TIFR/TH/99-34\\
IC/99/79\\
July 1999\\
hep-th/9907075
\end{flushright}
\begin{center}
\vspace{2 ex}
{\large{\bf D1/D5 Moduli in SCFT and Gauge Theory, and Hawking
Radiation}}\\
\vspace{3 ex}
Justin R. David$^{a,*}$, Gautam Mandal$^{a,\dagger}$, 
and Spenta R. Wadia$^{a,b,\ddagger}$ \\
$\,$\\
$^a${\sl Department of Theoretical Physics,
Tata Institute of Fundamental Research,}\\
{\sl Homi Bhabha Road, Mumbai 400 005, INDIA.} \\
$\,$ \\ 
$^b${\sl Abdus Salam International Centre for Theoretical
Physics,}
\\
{\sl Strada Costiera 11, Trieste 34014, ITALY.}

\vspace{10 ex}
\pretolerance=1000000
\bf ABSTRACT\\
\end{center}
\vspace{1 ex}

We construct marginal operators of the orbifold SCFT corresponding to
all twenty near-horizon moduli in supergravity, including operators
involving twist fields which correspond to the blowing up modes. We
identify the operators with the supergravity moduli in a 1-1 fashion
by inventing a global $SO(4)$ algebra in the SCFT. We analyze the
gauge dynamics of the D1/D5 system relevant to the splitting
$(Q_1,Q_5)\to (Q'_1,Q'_5)+ (Q''_1,Q''_5)$ with the help of a linear
sigma model.  We show in supergravity as well as in SCFT that the
absorption cross-section for minimal scalars is the same all over the
near-horizon moduli space.

\vfill
\sgap
\hrule
\vspace{0.5 ex}
\begin{flushleft}
\baselineskip 2ex
{\tenrm
e-mail:  justin,mandal,wadia@theory.tifr.res.in\\
$^*$ Address after August 31, 1999: 
Department of Physics, University of California,
Santa Barbara,\\ CA 93106, U.S.A. \\
$^\dagger$ Address after August 31, 1999: Theory Division,
CERN, CH 1211, Geneve 23, Switzerland.\\
$^\ddagger$ Jawaharlal Nehru Center for Advanced
Scientific Research. Bangalore 560012, INDIA.}
\end{flushleft}
\baselineskip 3.9ex

\section{Introduction}

The D1/D5 system has been crucial in understanding black hole physics
in string theory. Our current understanding of the microscopic
derivation of black hole entropy and Hawking radiation rests mainly on
this model \cite{StrVaf,CalMal,MalSus,DhaManWad,DasMat,MalStr}.  The
D1/D5 system also provides a concrete realization of holography in the
near horizon limit
\cite{Maldacena,MalStr98,Martinec,GivKutSei,deBoer,DavManWad,BalKraLaw}.
Unlike in other examples of $AdS_{d+1}/CFT_{d}$ duality, this system
provides an example where both sides of the duality are well
tractable.

The D1/D5 system is constructed as follows. Consider type IIB string
theory compactified on $T^4$ (one could also consider $K_3$ ). Let us
assume that the coordinates of the compactified directions are
$x^{6,7,8,9}$. Let us consider $Q_5$ D5-branes extending along
$x^{5,6,7,8,9}$ and wrap four of the directions along the $T^4$.  Also
let us consider $Q_1$ D1-branes extending along the coordinate
$x^5$. This leaves us with a black string in six dimensions carrying
electric as well as magnetic charge under the Ramond-Ramond field
$C^{(2)} \equiv B'$. The much studied example of the five dimensional
black hole solution is obtained by compactifying the $x^5$ coordinate
and introducing Kaluza-Klein momentum along this direction
\cite{CalMal}. The complete specification of the D1/D5 system includes
various moduli. Most of the study of the D1/D5 system so far has been
focused on the situation with no moduli. It is known from supergravity
that the D1/D5 system with no moduli turned on is marginally stable
with respect to decay of a subsystem consisting of $Q'_5$ D5 and
$Q'_1$ D1 branes. It has been observed recently \cite{SeiWit} that
such a decay in fact signals a singularity in the world volume gauge
theory associated with the origin of the Higgs branch.  The issue of
stability in supergravity in the context of the D1/D5 system has also
been discussed in \cite{Dijkgraaf,LarMar}.

The singularity mentioned above leads to a singular conformal field
theory and hence to a breakdown of string perturbation
theory. However, generic values of the supergravity moduli which do
not involve fragmentation into constituents are described by
well-defined conformal field theories and therefore string
perturbation theory makes sense.  In this paper, we would like to
understand the D1/D5 system at a generic point in the moduli space.
In particular, we address two issues: (a) What is the boundary SCFT
corresponding to the D1/D5 system at generic values of the moduli? (b)
To what extent do the various moduli affect the Hawking rate of the
minimal scalars computed from the microscopic SCFT?

The low energy description of the D1/D5 system is given by a ${\cal
N}=(4,4)$ SCFT on the moduli space of $Q_1$ instantons of a $U(Q_5)$
gauge theory on $T^4$.  This moduli space is conjectured to be a
resolution of the orbifold $(\widetilde{T}^4)^{Q_1Q_5}/S(Q_1Q_5)$
\cite{Vaf} which we shall denote by ${\cal M}$. $\widetilde{T}^4$ can
be distinct from the compactification torus $T^4$. The evidence for
this conjecture is mainly topological and is related to dualities
which map the black string corresponding to the D1/D5 system to a
perturbative string of Type IIB theory with $Q_1$ units of momentum
and $Q_5$ units of winding along the $x^5$ direction \cite{Sen}.  An
orbifold theory realized as a free field SCFT with identifications is
nonsingular as all correlations functions are finite. The realization
of the SCFT of the D1/D5 system as ${\cal N}=(4,4)$ theory on ${\cal
M}$ implies that we are at a generic point in the moduli space of the
D1/D5 system and {\em not at the singularity corresponding to
fragmentation}. In other words, the orbifold SCFT corresponds to a
bound state of $Q_1$ D1-branes and $Q_5$ D5-branes (henceforth denoted
as the $(Q_1,Q_5)$ bound state). Thus we use the free field
realization of ${\cal N}=(4,4)$ SCFT on the orbifold ${\cal M}$ and
its resolutions using the marginal operators of this theory to
describe the boundary SCFT corresponding to the D1/D5 system at
generic values of the moduli. We construct all the $20$ marginal
operators of this theory including the $4$ blow up modes
explicitly. We use symmetries, including a new global $SO(4)$ algebra,
to identify the marginal operators with their corresponding
supergravity moduli.

Once we have found the four marginal operators corresponding to the
blow up modes, we address the question how to understand their origin
in the gauge theory of the D1/D5 system and also how to describe the
splitting of the $(Q_1,Q_5)$ bound state into subsystems $(Q'_1,Q'_5)$
and $(Q''_1, Q''_5)$ in the gauge theory language. We find that the
dynamics of this splitting is described by an effective $(4,4)$ $U(1)$
theory coupled to $Q'_1 Q''_5 + Q''_1 Q'_5$ hypermultiplets.  We show,
by an analysis of the D-term equations and the potential, that the
splitting is possible only when the Fayet-Iliopoulos terms and the
theta term of the effective gauge theory are zero. These, therefore,
should correspond to the four SCFT marginal operators corresponding to
the blow up modes.

We also address the question whether the Hawking rate of minimal
scalars of the five dimensional black hole at a generic point in the
moduli space of the D1/D5 system agrees with the SCFT calculation. We
find that the absorption cross-section and Hawking rate do not depend
on the moduli in supergravity, essentially because the minimal scalars
are coupled only to the Einstein metric which remains unchanged under
U-duality transformations generating the moduli. In the SCFT the
calculation of the absorption cross-section/Hawking radiation depends
on the two-point function of the corresponding operators, and
we show that turning on the exactly marginal perturbations
do not modify the cross-section or the rate.

This paper is organized as follows. In Section 2 we construct all the
marginal operators of the ${\cal N}=(4,4)$ SCFT on ${\cal M}$ and
identify their quantum numbers under the symmetries of the SCFT. In
Section 3 we make a one-to-one identification of the supergravity
moduli with the marginal operators in the SCFT. In Section 4 we
analyze the gauge dynamics of the D1/D5 system relevant for splitting
into subsystems.  In Section 5 we discuss absorption/Hawking radiation
of the minimal scalars from supergravity and SCFT in the presence of
moduli. Section 6 contains concluding remarks.

\section{The resolutions of the symmetric product}

In this section we will construct marginal operators of the ${\cal
N}=(4,4)$ SCFT on the symmetric product orbifold ${\cal M}$. We will
find the four operators which correspond to resolution of the orbifold
singularity.

The SCFT is described by the free Lagrangian
\be
\label{free}
S = \frac{1}{2} \int d^2 z\; \left[\del
x^i_A \bar\del x_{i,A} + 
\psi_A^i(z) \bar\del \psi^i_A(z) + 
\widetilde\psi^i_A(\bar z) \del \widetilde \psi^i_A(\bar z) 
 \right]
\ee
Here $i$ runs over the $\widetilde{ T^4}$ coordinates
1,2,3,4 and $A=1,2,\ldots Q_1Q_5$ labels various copies
of the four-torus. 

In order to organize the fermions according to doublets of
$SU(2)_R \times \widetilde {SU(2)_R}$, we introduce the
following notations:

$\Psi_A(z)$ denotes the row vector
of fermions
\be
\Psi_A (z) \equiv (\Psi_A^1(z), \Psi_A^2(z)) \equiv 
\sqrt{1/2} (\psi_A^1(z) +
i\psi_A^2(z), \psi_A^3(z) + i\psi_A^4(z))
\ee
$\Psi_A^\dagger(z)$
denotes the column vector 
\be {\Psi_A}^\dagger(z) =
\left(
\begin{array}{c}
{\Psi_A}^{\dagger 1}(z) \\
{\Psi_A}^{\dagger 2}(z) \\
\end{array}
\right)
= \sqrt{\frac{1}{2}}
\left(
\begin{array}{c}
\psi_A^1(z) 
- i \psi_A^2(z) \\
\psi_A^3(z) 
- i \psi_A^4z) \\
\end{array}
\right)
\ee
Similarly $\widetilde \Psi_A(\bar z)$ will denote the antiholomorphic
counterparts of the above fermions. (See appendix A for 
more details.) 

\subsection{The untwisted sector}
Let us first focus on the operators constructed from the
untwisted sector. The operators of lowest conformal weight 
are
\bea 
\label{chiral}
\Psi^1_A(z) \widetilde{\Psi}^1_A(\bar{z})  \;&\;&\;  
\Psi^1_A(z)\widetilde{\Psi}^{2\dagger}_A(\bar z) \\   \nonumber
\Psi^{2\dagger}_A(z)\widetilde{\Psi}^1_A(\bar z )  \;&\;&\;
\Psi^{2\dagger}_A(z)\widetilde{\Psi}^{2\dagger}_A(\bar z) 
\eea
where summation over $A$ is implied. These four operators have conformal
dimension $(h, \bar{h})=(1/2, 1/2)$ and
$(j_R^3, \widetilde{j}_R^3)= (1/2, 1/2)$ under
the R-symmetry  $SU(2)_R\times \widetilde{SU(2)}_R$
(see Appendix A). Since $(h, \bar{h})=(j_R^3, \widetilde{j}_R^3)$,
 these operators are chiral primaries and
have non-singular operator product expansions (OPE) with the
supersymmetry currents 
$G^1(z), G^{2\dagger}(z), \widetilde{G}^1(\bar z),
\widetilde{G}^{2\dagger}(\bar z)$ 
(defined in Appendix A).  These properties indicate
that they belong to the bottom component of the short multiplet $(\bf
2, \bf 2)_S$ (See Appendix B for details).  Each of the four chiral
primaries gives rise to four top components of the short multiplet
$(\bf 2, \bf 2)_S$. They are given by the leading pole
($(z-w)^{-1} (\bar z - \bar w)^{-1}$) in the
OPE's
\bea
\label{OPE}
G^2(z)G^2(\bar z){\cal P} (w, \bar w)
\; &\;&\; 
G^2(z)\widetilde{G}^{1\dagger}(\bar z){\cal P} (w, \bar w)
\\ \nonumber
G^{1\dagger}(z) \widetilde{G}^2(\bar z){\cal P}(w, \bar w) \; &\;& \;
G^{1\dagger}(z)\widetilde{G}^{1\dagger} 
(\bar z){\cal P} (w, \bar w)
\eea
where ${\cal P}$ stands for any of the four chiral primaries in
\eq{chiral}. From the superconformal algebra it is easily seen that
the top components constructed above have weights $(1,1)$ and
transform as $(\bf 1, \bf 1 )$ under $SU(2)_R\times
\widetilde{SU(2)}_R$.  The OPE's \eq{OPE}\ can be easily evaluated. We
find that the $16$ top components of the $4 (\bf 2, \bf 2)_S$ short
multiplets are $\del x_A^i \bar{\del} x_A^j$.

We classify the above operators belonging to the top component
according to representations of (a) the $SO(4)_I$ rotational symmetry
(Appendix A) 
of the $\widetilde{T}^4$, (The four torus $\tilde{T}^4$ breaks this
symmetry but we assume the target space is $R^4$ for the
classification of states)
(b) $R$ symmetry of the SCFT and (c) the
conformal weights. As all of these operators belong to the top
component of $(\bf 2, \bf 2 )_{\bf S}$ the only property which
distinguishes them is the representation under $SO(4)_I$. 
The quantum
numbers of these operators under the various symmetries are
\bea
\label{untwist_operator}
\begin{array}{lccc}
\mbox{Operator}&SU(2)_I\times \widetilde{SU(2)}_I&
SU(2)_R\times\widetilde{SU(2)}_R& (h, \bar{h}) \\
\del x^{ \{ i }_A(z) \bar{\del}x^{ j\} }_A (\bar z) -
\frac{1}{4}\delta^{ij}
\del x^k_A(z) \bar{\del}x^k_A (\bar z) &(\bf 3, \bf 3) & (\bf 1,\bf 1) 
& (1, 1) \\  
\del x^{[i}_A(z) \bar{\del}x^{j]}_A (\bar z) & (\bf 3, \bf 1) +
(\bf 1, \bf 3) & (\bf 1, \bf 1)& (1,1) \\ 
\del x^i_A(z) \bar{\del}x^i_A (\bar z) &(\bf 1, \bf 1)& (\bf 1, \bf 1)
& (1,1)
\end{array}
\eea
Therefore we have $16$ marginal operators from the untwisted sector.
As these are top components they can be added to the free SCFT as
perturbations without violating the ${\cal N}=(4,4)$ supersymmetry.

\subsection{$Z_2$ twists.}

We now construct the marginal operators from the various twisted
sectors of the orbifold SCFT.  The twist fields of the SCFT on the
orbifold ${\cal M}$ are labeled by the conjugacy classes of the
symmetric group $S(Q_1 Q_5)$ \cite{VafWit,DijMooVerVer}. The conjugacy
classes consist of cyclic groups of various lengths. The various
conjugacy classes and the multiplicity in which they occur in
$S(Q_1Q_5)$ can be found from the solutions of the equation
\be
\sum nN_n = Q_1 Q_5
\ee
where $n$ is the length of the cycle and $N_n$ is the multiplicity of
the cycle. Consider the simplest nontrivial conjugacy class which is
given by $N_1 = Q_1 Q_5 -2, N_2 = 1$ and the rest of $N_n =0$.  
A representative element of this class is 
\be
\label{group_element}
(X_1\rightarrow X_2, \; X_2\rightarrow X_1), 
\; X_3\rightarrow X_3 , \; \ldots ,
\; X_{Q_1Q_5} \rightarrow X_{Q_1Q_5}
\ee
Here the $X_A$'s are related to the $x_A$'s appearing in 
the action \eq{free} by \eq{defn} in  Appendix A.

To exhibit the singularity of this group action we go over to the
following new coordinates
\be
X_{cm} = X_1+ X_2 \;\; \mbox{and}\;\; \phi = X_1-X_2
\ee
Under the group action \eq{group_element}\ 
$X_{cm}$ is invariant and $\phi
\rightarrow -\phi$. Thus the singularity is {\em locally} of the type
$R^4/Z_2$. The bosonic twist operators for this orbifold singularity
are given by following OPE's
\cite{DixFriMarShe}
\bea
\del \phi^1 (z) \sigma^1(w, \bar{w} ) = 
\frac{ \tau^1(w, \bar w ) }{ (z-w)^{1/2} }  \; &\;& \;
\del {\phi}^{1\dagger} (z) \sigma^1(w, \bar{w} ) = 
\frac{ \tau'^1(w, \bar w ) }{ (z-w)^{1/2} }  \\   \nonumber
\del \phi^2 (z) \sigma^2(w, \bar{w} ) =
\frac{ \tau^2(w, \bar w ) }{ (z-w)^{1/2} }  \; &\;& \;
\del {\phi}^{2\dagger} (z) \sigma^2(w, \bar{w} ) =
\frac{ \tau'^2(w, \bar w ) }{ (z-w)^{1/2} }  \\  \nonumber
\bar{\del} \widetilde{\phi}^1 (\bar{z}) \sigma^1(w, \bar{w} ) = 
\frac{ \widetilde\tau'^1(w, \bar w ) }{ (\bar z-\bar w)^{1/2} }  \; &\;&
\;
\bar{\del} \widetilde{\phi}^{1\dagger} (\bar z) \sigma^1(w, \bar{w} ) = 
\frac{ \widetilde\tau^1(w \bar w ) }{ (\bar z-\bar w)^{1/2} } 
\\   \nonumber
\bar{\del} \widetilde{\phi}^2 (\bar z) \sigma^2(w, \bar{w} ) =
\frac{ \widetilde\tau'^2(w, \bar w ) }{ (\bar z-\bar w)^{1/2} }  \; &\;&
\;
\bar{\del} \widetilde{\phi}^{2\dagger} (\bar z ) \sigma^2(w, \bar{w} ) =
\frac{ \widetilde\tau^2(w, \bar w ) }{ (\bar z-\bar w)^{1/2} }  
\eea
The $\tau$'s are excited twist operators.
The fermionic twists are constructed from bosonized currents defined
by
\bea
\chi^1(z) = e^{iH^1(z)} \; &\;& \; \chi^{1\dagger}(z) = e^{-iH^1(z)} \\
\nonumber
\chi^2(z) = e^{iH^2(z)} \; &\;& \; \chi^{2\dagger}(z) = e^{-iH^2(z)} \\
\nonumber
\eea
Where the $\chi$'s, defined as
$\Psi_1 - \Psi_2$, are the superpartners of the bosons $\phi$.

{}From the above we construct the supersymmetric twist fields which act
both on fermions and bosons as follows:
\bea
\label{defSigma}
\Sigma^{(\frac{1}{2}, \, \frac{1}{2})}_{(12)} = \sigma^1(z,\bar z)
\sigma^2 (z,\bar z)
e^{iH^1(z)/2} e^{-iH^2(z)/2} 
e^{i\widetilde{H}^1(\bar z )/2} e^{-i\widetilde{H}^2(\bar z )/2}
\\  \nonumber
\Sigma^{(\frac{1}{2},\,  -\frac{1}{2})}_{(12)} = \sigma^1(z,\bar z)
\sigma^2 (z,\bar z)
e^{iH^1(z)/2} e^{-iH^2(z)/2} 
e^{- i\widetilde{H}^1(\bar z )/2} e^{i\widetilde{H}^2(\bar z )/2} 
\\  \nonumber
\Sigma^{(-\frac{1}{2}, \,  \frac{1}{2})}_{(12)} = \sigma^1(z,\bar z)
\sigma^2 (z,\bar z)
e^{-iH^1(z)/2} e^{+iH^2(z)/2} 
e^{i\widetilde{H}^1(\bar z )/2} e^{-i\widetilde{H}^2(\bar z )/2} 
\\   \nonumber
\Sigma^{(-\frac{1}{2}, \,  -\frac{1}{2})}_{(12)} = 
\sigma^1(z,\bar z) \sigma^2 (z,\bar z)
e^{-iH^1(z)/2} e^{+iH^2(z)/2} 
e^{-i\widetilde{H}^1(\bar z )/2} e^{+i\widetilde{H}^2(\bar z )/2} 
\\  \nonumber
\eea
The subscript $(12)$ refers to the fact that these twist operators were
constructed for the representative
group element \eq{group_element}\ which exchanges the $1$ and
$2$ labels of the coordinates of $\widetilde{T}^4$. 
The superscript stands
for the $(j^3_R, \widetilde{j}^3_R)$ quantum numbers.
The twist operators for the orbifold ${\cal M}$ belonging to the
conjugacy class under consideration is obtained by summing over these
$Z_2$ twist operators for all representative elements of this class.
\be
\Sigma^{(\frac{1}{2}, \,  \frac{1}{2})} =
\sum_{i=1}^{Q_1Q_5} \sum_{j=1, j\neq i}^{Q_1Q_5}
 \Sigma^{(\frac{1}{2}, \,  \frac{1}{2})}_{(ij)}
\ee
We can define the rest of the twist operators for the orbifold in a
similar manner. The conformal dimensions of these operators are
$(1/2,1/2)$. They transform as $(\bf 2 , \bf 2)$ under the $SU(2)_R
\times \widetilde{SU(2)}_R$ symmetry of the SCFT. They belong to the
bottom component of the short multiplet $(\bf 2, \bf 2)_S$. The
operator $\Sigma^{(\frac{1}{2}, \, \frac{1}{2})}$ is a chiral primary.
As before the $4$ top components of this short multiplet,
which we denote by
\bea
T^{(\frac{1}{2}, \, \frac{1}{2})}, \;\; T^{(\frac{1}{2}, \,
-\frac{1}{2})} \\ \nonumber
T^{(-\frac{1}{2}, \, \frac{1}{2})}, \;\;
T^{(-\frac{1}{2}, \, -\frac{1}{2})} 
\eea 
are given  by the leading pole in the following OPE's respectively
\bea 
\label{raising}
G^2(z)\widetilde{G}^2(\bar z)\Sigma^
{(\frac{1}{2}, \,  \frac{1}{2})} (w, \bar w), \;\;
G^2(z) \widetilde{G}^{1\dagger}(\bar z)\Sigma^{(\frac{1}{2}, \,  
\frac{1}{2})} (w, \bar w),  \\  \nonumber
 G^{1\dagger}(z) \widetilde{G}^2(\bar z)
 \Sigma^{(\frac{1}{2}, \, \frac{1}{2})} (w, \bar w), \;\;
G^{1\dagger}(z)\widetilde{G}^{1\dagger}
(\bar z)\Sigma^{(\frac{1}{2}, \,\frac{1}{2})} (w, \bar w)
\eea
These are the $4$ blow up modes of the $R^4/Z_2$ singularity
\cite{CveDix}\ and they have conformal weight $(1,1)$%
\footnote{Relevance of $Z_2$ twist operators
to the marginal deformations of the SCFT has earlier
been discussed in \cite{HasWad1,DijVerVer}}. They transform as
$(\bf 1 , \bf
1)$ under the $SU(2)_R \times \widetilde{SU(2)_R}$.  As before, since
these are top components of the short multiplet $(\bf 2 , \bf 2)_S$
they can be added to the free SCFT as perturbations without violating
the ${\cal N} = (4,4)$ supersymmetry of the SCFT. 
The various quantum numbers of
these operators are listed below.
\bea
\begin{array}{lccc}
\label{twist_operator}
\mbox{Operator} & (j^3, \widetilde{j}^3)_I & 
SU(2)_R\times \widetilde{SU(2)}_R &  (h, \bar{h}) \\
{\cal T}^1_{(1)}=T^{(\frac{1}{2}, \,  \frac{1}{2})} & 
( 0, 1) & (\bf 1, \bf 1) & (1,1) \\
{\cal T}^1_{(0)}= T^{(\frac{1}{2}, 
\,  -\frac{1}{2})}+ T^{(-\frac{1}{2}, \,  \frac{1}{2})} &  
(0,0) & (\bf 1, \bf 1) & (1,1) \\
{\cal T}^1_{(-1)}= T^{(-\frac{1}{2}, \,  -\frac{1}{2})} 
& (0,-1) & (\bf 1, \bf 1) & (1,1) \\
{\cal T}^0= T^{(-\frac{1}{2}, \, -\frac{1}{2})} - T^{(-\frac{1}{2}, 
\,  -\frac{1}{2})}
& (0, 0) & (\bf 1, \bf 1) & (1,1) 
\end{array}  
\eea
The first three operators of the above table can be organized as a
$(\bf 1, \bf 3)$ under $SU(2)_I\times\widetilde{SU(2)}_I$. We will
denote these $3$ operators as ${\cal T}^1$. The last
operator transforms as a scalar $(\bf 1, \bf 1)$ under 
$SU(2)_I\times\widetilde{SU(2)}_I$ and is denoted by ${\cal T}^0$. 
The simplest way of figuring out the $(j^3, \widetilde{j}^3)_I$
quantum numbers in the above
table is to note that
(a) the $\Sigma$-operators
of \eq{defSigma} are singlets under $SU(2)_I\times\widetilde{SU(2)}_I$,
as can be verified by computing the action on them of
the operators $I_1, I_2$ and $\widetilde{I}_1,
\widetilde{I}_2$, (b) the ${\cal T}$-operators are
obtained from $\Sigma$'s by the action of the supersymmetry
currents as in \eq{raising} and (c) the
quantum numbers of the supersymmetry
currents under $I_1, I_2$ and $\widetilde{I}_1,
\widetilde{I}_2$ are given by \eq{so4_on_g}. 

\subsection{Higher twists}

We now show that the twist operators corresponding to any other
conjugacy class of $S(Q_1 Q_5)$ are irrelevant. Consider the class
with $N_1= Q_1Q_5-3, N_3 =1$ and the rest of $N_n=0$. A representative
element of this class is 
\be
\label{z3_element}
(X_1\rightarrow X_2, X_2 \rightarrow X_3, X_3 \rightarrow X_1), 
\; X_4\rightarrow X _4, \ldots ,
\; X_{Q_1Q_5}\rightarrow X_{Q_1Q_5}. 
\ee
To make the action of this group element transparent we diagonalize
the group action as follows. 
\bea
\left(
\begin{array}{c}
\phi_1 \\ \phi_2 \\ \phi_3
\end{array} \right)
=
\left(
\begin{array}{ccc}
1 & 1 & 1 \\
1 & \omega & \omega^2 \\
1 & \omega^2 & \omega^4 
\end{array}
\right)
\left(
\begin{array}{c}
X_1 \\ X_2 \\  X_3
\end{array}
\right)
\eea
where $\omega = \exp(2\pi i/3)$.
These new coordinates are identified under the group action 
\eq{z3_element} $\phi_1
\rightarrow \phi_1$, $\phi_2 \rightarrow \omega^2 \phi_2$  and
$\phi_3 \rightarrow \omega \phi_3$. 
These identifications are locally characteristic of the
orbifold
\be
R^4\times R^4/\omega \times R^4/\omega^2
\ee
The dimension of the supersymmetric twist operator which twists the
coordinates by a phase $e^{2\pi i k /N}$ in $2$ complex dimensions is
$h(k,N)= k/N$\cite{DixFriMarShe}. The twist operator which implements the
action of the group element \eq{z3_element} combines the
supersymmetric twist operators acting on $\phi_2$ and $\phi_3$
and therefore has total dimension  
\be
h =h(1,3) + h(2,3) = 
1/3+ 2/3 =1 
\ee 
It is
the superpartners of these which could be candidates for the blow up
modes. However, these have weight $3/2$, These operators are therefore
irrelevant.

For the class $N_1= Q_1Q_5 -k$ , $N_k =k$ and the rest of $N_n =0$,
the total dimension of the twist operator is 
\be
h = \sum_{i=1}^{k-1} h(i,k) = (k-1)/2
\ee 
Its superpartner has dimension $k/2$. Now it is easy to see that all
conjugacy classes other than the exchange of $2$ elements give
rise to irrelevant twist operators. Thus the orbifold ${\cal M}$ is
resolved by the $4$ blow up modes corresponding to the conjugacy class
represented by \eq{group_element}. We have thus
identified the $20$ marginal operators of the ${\cal N}=(4,4)$ SCFT on
$\widetilde{T}^4$. They are all top components of the $5(\bf 2 ,\bf
2)_S$ short multiplets.

\section{The supergravity moduli and the resolutions of the orbifold.}

In this section we find the massless scalar fields which couple to the
$4$ blow up modes of the orbifold ${\cal M}$.  Type IIB supergravity
compactified on $T^4$ has 25 scalars. There are $10$ scalars $h_{ij}$
which arise from compactification of the metric.  $i, j, k \ldots $
stands for the directions of $T^4$. There are $6$ scalars $b_{ij}$
which arise from the Neveu-Schwarz $B$-field and similarly there are
$6$ scalars $b'_{ij}$ from the Ramond-Ramond $B'$-field. The remaining
3 scalars are the dilaton $\phi$, the Ramond-Ramond scalar $\chi$ and
the Ramond-Ramond $4$-form $C_{6789}$. These scalars parameterize the
coset $SO(5,5)/(SO(5)\times SO(5))$.  The near horizon limit of the
D1/D5 system is $AdS_3\times S^3\times T^4$ \cite{MalStr}. 
In this geometry $5$ of
the $25$ scalars become massive.  They are the $h_{ii}$ (the trace of
the metric of $T^4$ which is proportional to the volume of $T^4$), the
$3$ components of the anti-self dual part of the Neveu-Schwarz
$B$-field $b_{ij}^-$ and a linear combination of the Ramond-Ramond
scalar and the $4$-form \cite{SeiWit}.  The massless scalars in the
near horizon geometry parameterize the coset $SO(5,4)/SO(5)\times
SO(4)$ \cite{GivKutSei}.

The near horizon symmetries form the supergroup $SU(1,1|2)\times
SU(1,1|2)$.  This is the global part of the ${\cal N}=(4,4)$
superconformal algebra.  We can classify \cite{deBoer,DavManWad} all
the massless supergravity fields of Type IIB supergravity on
$AdS_3\times S^3\times T^4$ ignoring the Kaluza-Klein modes on $T^4$
according to the short multiplets of the supergroup $SU(1,1|2)\times
SU(1,1|2)$. The massless fields of the supergravity fall into the top
component of the $5(\bf 2, \bf 2 )_S$ short multiplet.  We further
classify these fields according to the representations of the
$SO(4)_I$, the rotations of the $x^{6,7,8,9}$ directions.  This
is not a symmetry of the supergravity as it is compactified on $T^4$,
but it can be used to classify states. The quantum number of the
massless supergravity fields are listed below.
\bea
\label{sugra_fields}
\begin{array}{lccc}
\mbox{Field} & SU(2)_I\times \widetilde{SU(2)_I} 
& SU(2)_E\times \widetilde{SU(2)}_E &
\mbox{Mass} \\
h_{ij } -\frac{1}{4}
\delta_{ij} h_{kk} & (\bf 3, \bf 3) & (\bf 1, \bf 1) & 0 \\
b'_{ij} & (\bf 3, \bf 1) + (\bf 1, \bf 3) &(\bf 1, \bf 1) &  0 \\
\phi &  (\bf 1, \bf 1) &(\bf 1, \bf 1) &  0 \\
a_1 \chi + a_2 C_{6789}&  (\bf 1, \bf 1) &(\bf 1, \bf 1) &  0 \\
b^+_{ij} &  (\bf 1, \bf 3) &(\bf 1, \bf 1) & 0 
\end{array}
\eea
The linear combination appearing on the fourth line is the
one that remains massless in the near-horizon limit.
The $SU(2)_E\times \widetilde{SU(2)}_E$ 
stands for the $SO(4)$ isometries of the
$S^3$. All the above fields are s-waves of scalars in the near horizon
geometry. 

We would like to match the twenty supergravity moduli appearing
in \eq{sugra_fields} with the twenty marginal operators 
appearing in \eq{untwist_operator} and \eq{twist_operator} 
by comparing their symmetry properties under the AdS/CFT correspondence
\cite{DavManWad}. 

\noindent
The symmetries, or equivalently quantum numbers, to be compared under
the  AdS/CFT correspondence
\cite{Maldacena,Witten98,GubKlePol,MalStr98} are as follows:

(a) The isometries of the supergravity are identified with the global
symmetries of the superconformal field theory. For the $AdS_3$ case
the symmetries form the supergroup $SU(1,1|2)\times SU(1,1|2)$.  
The identification of this supergroup with the
global part of the ${\cal N}= (4,4)$ superalgebra leads
to the following mass-dimension relation
\be
h+\bar{h}= 1+ \sqrt{1+m^2}
\ee
where $m$ is the mass of the bulk field and $(h,\bar h)$
are the dimensions of the SCFT operator. Since in 
our case the SCFT operators are marginal and the supergravity
fields are massless, the mass-dimension relation is obviously
satisfied. 

(b) The $SU(2)_E\times \widetilde{SU(2)}_E$
quantum number of the bulk supergravity field corresponds to the
$SU(2)_R\times SU(2)_R$ quantum number of the boundary operator.
By an inspection of column three of the tables in
\eq{untwist_operator}, \eq{twist_operator} and \eq{sugra_fields}, 
we see that these quantum numbers
also match.
 
(c) The supersymmetry properties of the bulk fields and the boundary
operators tell us which component of the short multiplet they belong
to. Noting the fact that all the twenty bulk fields as
well as all the marginal operators mentioned above correspond to top
components of short multiplets, this property also matches.  

(d) The above symmetries alone do not distinguish between the twenty
operators or the twenty bulk fields. To further distinguish these
operators and the fields we identify the $SO(4)_I$ symmetry of the
directions $x^{6,7,8,9}$ with the $S0(4)_I$ of the SCFT. At the level
of classification of states this identification is reasonable though
these are not actual symmetries. Using the quantum numbers under this
group we obtain the following matching of the boundary operators and
the supergravity moduli.
\bea
\begin{array}{llc}
\mbox{Operator} & \mbox{Field}  &  
SU(2)_I\times \widetilde{SU(2)}_I \\
\del x^{ \{ i }_A(z) \bar{\del}x^{ j\} }_A (\bar z) -1/4\delta^{ij}
\del x^{k}_A \bar{\del}x^k_A
& h_{ij} -1/4\delta_{ij} h_{kk} & (\bf 3, \bf 3 ) \\
\del x^{[i}_A(z) \bar{\del}x^{j]}_A (\bar z) 
& b'_{ij}
& (\bf 3, \bf 1 ) +
(\bf 1, \bf 3)  \\ 
\del x^i_A(z) \bar{\del}x^i_A (\bar z) 
& \phi
&( \bf 1, \bf 1 ) \\
{\cal T}^1 & b^+_{ij} & (\bf 1, \bf 3 ) \\
{\cal T}^0 & a_1\chi + a_2C_{6789} & (\bf 1, \bf 1 )
\end{array}
\eea
Note that both the representations ${\bf (1,3)}$ and ${\bf (1,1)}$ occur
twice in the above table. This could give rise to a two-fold ambiguity
in identifying either ${\bf (1,3)}$ or ${\bf (1,1)}$ operators with
their corresponding bulk fields. The way we have resolved it here is
as follows.  The operators ${\cal T}^1$ and ${\cal T}^0$ correspond to
blow up modes of the orbifold, and since these are related to the
Fayet-Iliopoulos terms and the $\theta$-term in the gauge theory (see
Section 4), tuning these operators one can reach the singular SCFT
\cite{SeiWit} that corresponds to fragmentation of the D1/D5
system. In supergravity, similarly, it is only the moduli $b^+_{ij}$
and $a_1\chi + a_2C_{6789}$ which affect the stability of the D1/D5
system \cite{SeiWit,LarMar,Dijkgraaf}. As a result, it
is $b^+_{ij}$ (and not $b^{\prime +}_{ij}$) which
should correspond to the operator ${\cal T}^1$ and similarly
$a_1\chi + a_2C_{6789}$ should correspond to ${\cal T}^0$.

Thus, we arrive at a one-to-one identification between
operators of the SCFT and the supergravity moduli.

\section{The linear sigma model}

In this section we will analyze the gauge theory description of the
D1/D5 system.  We show that that the gauge theory  has four 
parameters which control the break up of the $(Q_1, Q_5)$ system to 
subsystems $(Q'_1, Q'_5)$ and $(Q''_1, Q''_5)$. These are the
Fayet-Iliopoulos D-terms and the theta term in the effective $U(1)$
$(4,4)$ linear sigma model of the relative coordinate between the
subsystems $(Q'_1, Q'_5)$ and $(Q''_1, Q''_5)$.
To motivate this 
we will review the linear  sigma model corresponding to the
of the $R^4/Z_2$ singularity. 

\subsection{The linear sigma model description of $R^4/Z_2$}

The linear sigma model is a $1+1$ description of the $R^4/Z_2$
singularity 
dimensional $U(1)$ gauge theory with $(4,4)$ supersymmetry \cite{Wit}. 
It has $2$
hypermultiplets charged under the $U(1)$. The scalar fields of the 
hypermultiplets can be
organized as doublets under the $SU(2)_R$ symmetry of the $(4,4)$
theory as
\bea
\chi_1=
\left(
\begin{array}{c}
A_1 \\ B_1^\dagger
\end{array}
\right)  \;\mbox{and} \;
\chi_2=
\left(
\begin{array}{c}
A_2 \\ B_2^\dagger
\end{array}
\right) 
\eea
The $A$'s have charge $+1$ and the $B$'s have charge $-1$ under the
$U(1)$. The vector multiplet has $4$ real 
scalars $\varphi_i$, $i=1,\ldots ,4$.
They do not transform under the $SU(2)_R$. One can 
include $4$ parameters
in this theory consistent with (4,4) supersymmetry. They are the $3$
Fayet-Iliopoulos terms and the theta term. 

Let us first investigate the hypermultiplet moduli space of this
theory with the $3$ Fayet-Iliopoulos terms and the theta term set to
zero. The Higgs phase of this theory is obtained by setting $\phi_i$
and the D-terms to zero. The D-term equations are
\bea
\label{dtermr4}
|A_1|^2 + |A_2|^2 -|B_1|^2 -|B_2|^2 &=&0 \\  \nonumber
A_1B_1 + A_2B_2 &=&0
\eea
The hypermultiplet moduli space is the space of solutions of the above
equations modded out by the $U(1)$ gauge symmetry. Counting the number
of degrees of freedom indicate that this space is $4$ dimensional. To
obtain the explicit form of this space it is convenient to introduce
the following gauge invariant variables
\bea
\label{dtermdef1}
M=A_1B_2 \; &\;& \; N= A_2B_1 \\ \nonumber
P= A_1B_1 &=& - A_2B_2 \\
\eea
These variables are not independent. Setting the D-terms equal to zero
and modding out the resulting space by $U(1)$ is equivalent to the
equation
\be
\label{surface}
P^2 +MN=0.
\ee
This homogeneous equation is an equation of the space $R^4/Z_2$. To see
this the solution of the above equation can be parametrized by $2$
complex numbers $(\zeta , \eta)$ such that
\be
\label{dtermdef2}
P= i\zeta \eta \;\; M=\zeta^2 \;\; N= \eta^2
\ee
Thus the point $(\zeta, \eta)$ and $(-\zeta, -\eta)$ are the same
point in the space of solutions of \eq{surface}. We have
shown that the hypermultiplet moduli space is $R^4/Z_2$. 

The above singularity at the origin of the moduli space 
is a geometric singularity in the hypermultiplet
moduli space. We now argue that this singularity is a genuine
singularity of the SCFT that the linear sigma model flows to in the
infrared. At the origin of the classical moduli space the Coulomb
branch meets the Higgs branch. In addition to the potential due to the
D-terms the linear sigma model contains the following term in the
superpotential\footnote{These terms can be understood from
the coupling $A_\mu A^\mu \chi^* \chi$ in six dimensions,
and recognizing that under dimensional reduction to two
dimensions $\varphi_i$'s appear from the components of $A_\mu$
in the compact directions.}
\be
V = 
(|A_1|^2 + |A_2|^2 + |B_1|^2 + |B_2|^2)(\varphi_1^2 + 
\varphi_2^2 +\varphi_3^2 +\varphi_4^2)
\ee
Thus at the origin of the hypermultiplet moduli space a flat direction
for the Coulomb branch opens up. The ground state at this point is not
normalizable due to the non-compactness of the Coulomb branch. This
renders the infrared SCFT singular.

This singularity can be avoided in two distinct ways. If one turns on
the Fayet-Iliopoulos D-terms, the D-term equations are modified to
\cite{Wit} 
\bea |A_1|^2 + |A_2|^2 -|B_1|^2 -|B_2|^2 &=&r_3 \\
\nonumber A_1B_1 + A_2B_2 &=&r_1+ir_2 
\eea 
Where $r_1,r_2,r_3$ are the
$3$ Fayet-Iliopoulos D-terms transforming as the adjoint of the
$SU(2)_R$. Now the origin is no more a solution of these equations and
the non-compactness of the Coulomb branch is avoided. In this case
wave-functions will have compact support all over the hypermultiplet
moduli space. This ensures that the infrared SCFT is
non-singular. Turning on the Fayet-Iliopoulos D-terms thus correspond
to the geometric resolution of the singularity. The resolved
space is known  to be \cite{Wit,Aspinwall} described by 
an Eguchi-Hanson metric in which $r_{1,2,3}$ parameterize
a shrinking two-cycle.

The second way to avoid the singularity in the SCFT is to turn on the
theta angle $\theta$. This induces a constant electric field in the
vacuum. This electric field is screened at any other point than the
origin in the hypermultiplet moduli space as the $U(1)$ gauge field is
massive with a mass proportional to the vacuum expectation value of the
hypers. At the origin the $U(1)$ field is not screened and thus it
contributes to the energy density of the vacuum. This energy is
proportional to $\theta^2$. Thus turning on the theta term lifts the
flat directions of the Coulomb branch. This ensures that the
corresponding infrared SCFT is well defined though the hypermultiplet
moduli space remains geometrically singular. In terms of the
Eguchi-Hanson space, the $\theta$-term corresponds to a flux of the
antisymmetric tensor through the two-cycle mentioned above.

The $(4,4)$ SCFT on $R^4/Z_2$ at the orbifold point is well defined.
Since the orbifold has a geometric singularity but the SCFT is
non-singular it must correspond to the linear sigma model with a
finite value of $\theta$ and the Fayet-Iliopoulos D-terms set to
zero.  Deformations of the $R^4/Z_2$ orbifold by its $4$ blow up modes
correspond to changes in the Fayet-Iliopoulos D-terms and theta term
of the linear sigma model%
\footnote{If we identify the $SU(2)_R$ of the linear sigma-model
with $\widetilde{SU(2)}_I$ of the orbifold SCFT, then the
Fayet-Iliopoulos parameters will correspond to ${\cal T}^1$
and the $\theta$-term to ${\cal T}^0$. 
This is consistent with Witten's observation \cite{Wit_cft}
that $SO(4)_E$ symmetry  of the linear sigma-model (one
that rotates the $\phi_i$'s) corresponds to the $SU(2)_R$
of the orbifold SCFT.} 
The global description of the moduli of a
${\cal N} =(4,4)$ SCFT on a resolved $R^4/Z_2$ orbifold is provided by
the linear sigma model.  In conclusion let us describe this linear
sigma model in terms of the gauge theory of D-branes. The theory
described above arises on a single D1-brane in presence of $2$
D5-branes. The singularity at the at the point $r_1, r_2, r_3, \theta
=0$ is due to noncompactness of the flat direction of the Coulomb
branch.  Thus it corresponds to the physical situation of the D1-brane
leaving the D5-branes.

\subsection{The gauge theory description of the moduli of the D1/D5
system}

As we have seen in Section 2 the resolutions of the ${\cal N}=(4,4)$
SCFT on ${\cal M}$ is described by 4 marginal operators which were
identified in the last subsection with the Fayet-Iliopoulos D-terms
and the theta term of the linear sigma model description of the
$R^4/Z_2$ singularity. We want to now indicate how these four
parameters would make their appearance in the gauge theory description
of the full D1/D5 system.

The gauge theory relevant for understanding the low energy degrees of
freedom of the D1/D5 system is a $1+1$ dimensional $(4,4)$
supersymmetric gauge theory with gauge group $U(Q_1)\times U(Q_5)$
\cite{Malthesis,HasWad}.
The matter content of this theory consists of hypermultiplets in the
adjoint representation of the two gauge groups $U(Q_1)$ and  $U(Q_5)$
They can be arranged as doublets under $SU(2)_R$ of the gauge theory
as
\be
N^{(1)}_{a\bar{a}} =
\left(
\begin{array}{c}
Y_{4(a\bar{a})}^{(1)} + iY_{3(a\bar{a)}}^{(1)} \\
Y_{2(a\bar{a})}^{(1)} - i Y_{1(a\bar{a)}}^{(1)} 
\end{array}
\right)
\quad \mbox{and}  \quad
N^{(5)}=
\left(
\begin{array}{c}
Y_{4(b\bar{b})}^{(5)} + iY_{3(b\bar{b})}^{(5)} \\
Y_{2(b\bar{b})}^{(5)} - i Y_{1(b\bar{b})}^{(5)} 
\end{array}
\right)
\ee
where $a,\bar{a}$ runs from $1,\ldots ,Q_1$ and $b,\bar{b}$ runs from
$1,\dots ,Q_5$. 
The $N^{(1)}$ transform as adjoints of $U(Q_1)$ and the $N^{(5)}$
transform as adjoints of $U(Q_5)$.  
$N^{(1)}$ corresponds to massless
excitations of open strings joining the D1-branes among themselves and
$N^{(5)}$ corresponds  to massless excitations of open strings 
joining D5-branes among themselves. This is clear from the expression
for the $N$'s in terms of the $Y$'s. The $Y$'s appear from the
components of the gauge field $A_\mu$ along the compact directions of
the four torus $T^4$.
The gauge theory also has hypermultiplets 
transforming as bi-fundamentals of $U(Q_1)\times \overline{U(Q_5)}$. 
These hypermultiplets can be arranged as doublets of the $SU(2)_R$
symmetry of the theory as
\be
\chi_{a \bar b}=
\left(
\begin{array}{c}
A_{a \bar b } \\B^\dagger_{a \bar b }
\end{array}
\right)
\ee
The hypermultiplets arise from massless excitations of open strings
joining the D1-branes and the D5-branes.
This gauge
theory was analyzed in detail in \cite{HasWad}. Motivated by the
D-brane description of the $R^4/Z_2$ singularity we look for the
degrees of freedom characterizing the break up of $(Q_1, Q_5)$ system
to $(Q'_1, Q'_5)$ and $(Q''_1, Q''_5)$. Physically the relevant degree
of freedom describing this process is the relative coordinate between
the centre of mass of the $(Q'_1, Q'_5)$ system and the $(Q''_1,
Q''_5)$. We will describe the effective theory of this degree of
freedom below.

For the bound state $(Q_1, Q_5)$ the hypermultiplets the $\chi_{a,
\bar{b}}$ are charged under the relative $U(1)$ of $U(Q_1)\times
U(Q_5)$, that is under the gauge field $A_\mu = Tr_{U(Q_1)}( A^{a \bar
a}_\mu) - Tr_{U(Q_5)}(A^{b \bar b})$.  The gauge multiplet
corresponding to the relative $U(1)$ corresponds to the degree of
freedom of the relative coordinate between the centre of mass of the
collection of $Q_1$ D1-branes and $Q_5$ D5-branes. At a generic point
of the Higgs phase, all the $\chi_{a\bar{b}}$'s have expectation
values, thus making this degree of freedom becomes massive. This is
consistent with the fact that we are looking at the bound state $(Q_1,
Q_5)$.

Consider the break up of the $(Q_1, Q_5)$ bound state to 
the bound states $(Q'_1, Q'_5)$ and $(Q''_1, Q''_5)$. To find out the
charges of the hypermultiplets under the various $U(1)$, we will
organize the hypers as
\bea
\left(
\begin{array}{cc}
\chi_{a'b'} &  \chi_{a'b''}    \\ 
\chi_{a''b'} & \chi_{a''b''}
\end{array}
\right), \quad
\left(
\begin{array}{cc}
Y_{i(a'\bar{a}')}^{(1)} & Y_{i(a'\bar{a}'')}^{(1)}  \\
Y_{i(a''\bar{a}')}^{(1)} & Y_{i(a''\bar{a}'')}^{(1)}
\end{array}
\right)
\quad \mbox{and} \quad
\left(
\begin{array}{cc}
Y_{i(b'\bar{b}')}^{(5)} & Y_{i(b'\bar{b}'')}^{(5)}  \\
Y_{i(b''\bar{b}')}^{(5)} & Y_{i(b''\bar{b}'')}^{(5)}
\end{array}
\right)
\eea
where $a',\bar{a}'$ runs from $1,\ldots , Q'_1$, $b',\bar{b}'$  
from $1,\dots ,Q'_5$,
$a''\bar{a}''$ from $1, \ldots Q''_1$ and 
$b'',\bar{b}''$ from $1\ldots ,Q''_5$. 
We organize the scalars of the vector multiplet corresponding to the
gauge group $U(Q_1)$ and $U(Q_5)$ as
\bea
\left(
\begin{array}{cc}
\phi_{i}^{^(1)a'\bar{a}'} & \phi_{i}^{(1)a'\bar{a}''} \\
\phi_{i}^{^(1)a''\bar{a}'}& \phi_{i}^{(2)a'\bar{a}''}
\end{array}
\right)
\quad
\mbox{and}
\quad
\left(
\begin{array}{cc}
\phi_{i}^{^(5)b'\bar{b}'} & \phi_{i}^{(5)b'\bar{b}''} \\
\phi_{i}^{^(5)b''\bar{b}'}& \phi_{i}^{(5)b'\bar{b}''}
\end{array}
\right)
\eea
where $i=1,2,3,4$. 

Let us call the the $U(1)$ gauge fields (traces) of 
$\, U(Q'_1),\, U(Q'_5),
U(Q''_1),\, U(Q''_5)\, $ as $\, A'_1,\,  A'_5, \, A''_1, \, A''_5\, $ 
respectively.  We
will also use the notation $A'_\pm \equiv A'_1 \pm A'_5$ and $A''_\pm
\equiv A''_1 \pm A''_5$.

As we are interested in the bound states $(Q'_1, Q'_5)$ and $(Q''_1,
Q''_5)$, in what follows we will work with a specific classical
background in which we give {\em vev}'s to the block-diagonal hypers
$\chi_{a'b'},\chi_{a''b''}, Y^{(1)}_{i( a' \bar a')},Y^{(5)}_{i(b'
\bar b')},Y^{(1)}_{i( a'' \bar a'')}$ and $Y^{(5)}_{i(b'' \bar b'')}$.
These {\em vev}'s are chosen so that the classical background
satisfies the D-term equations \cite{HasWad}.

The {\em vev}'s of the $\chi$'s  render the fields
$A'_-$ and $A''_-$ massive with a mass proportional to {\em vev}'s. In
the low energy effective Lagrangian these gauge fields can therefore
be neglected.  In the following we will focus on the $U(1)$ gauge
field $A_r=1/2(A'_+ - A''_+)$ which does not get mass from the above
{\em vev}'s. The gauge multiplet corresponding to $A_r$ contains four
real scalars denoted below by $\varphi_i$. These represent the
relative coordinate between the centre of mass of the $(Q'_1, Q'_5)$
and the $(Q''_1, Q''_5)$ bound states. We will be interested in the
question whether the $\varphi_i$'s remain massless or otherwise. The
massless case would correspond to a non-compact Coulomb branch and
eventual singularity of the SCFT.

In order to address the above question we need to find the low energy
degrees of freedom which couple to the gauge multiplet corresponding
to $A_r$.  

The fields charged under $A_r$ are the hypermultiplets 
$\, \chi_{a'\bar
b''},\;  \chi_{a''\bar b'}, \; Y^{(1)}_{i(a'\bar a'')}, \;
Y^{(1)}_{i(a''\bar a')},$ \\
$Y^{(5)}_{i(b'\bar b'')},\,  Y^{(5)}_{i(b''\bar b')}\, $ 
and the vector multiplets
$\phi^{(1)a'\bar a''}_{i}, \phi^{(1){a''\bar a'}}_i, 
\phi^{(5)b'\bar
b''}_{i}, \phi^{(5)b''\bar
b'}_i$. In order to 
find out which of these are massless, we
look at the following terms in the Lagrangian of
 $U(Q_1)\times U(Q_5)$ gauge theory:
\bea
\label{lagrangian}
L &=& L_1 + L_2 + L_3 + L_4\\  \nonumber
L_1&=& 
\chi_{a_1\bar{b}_1}^*\phi_i^{(1)a_2\bar{a}_1*}
\phi_i^{(1)a_2\bar{a}_3}\chi_{a_3\bar{b_1}} \\ \nonumber
L_2&=&
\chi_{a_1\bar{b_1}}^*\phi_i^{(5)b_1\bar{b}_2*}
\phi_i^{(5)b_3\bar{b}_2}\chi_{a_1\bar{b}_3} \\  \nonumber
L_3&=&Tr([Y^{(1)}_i,Y^{(1)}_j][Y^{(1)}_i, Y^{(1)}_j]) \\  \nonumber
L_4&=&Tr([Y^{(5)}_i,Y^{(5)}_j][Y^{(5)}_i, Y^{(5)}_j]) \\   \nonumber
\eea
where the  $a_i$'s run from $1,\ldots,Q_1$ 
and the $b_i$'s run form $1,\ldots, Q_5$. 
The terms $L_1$ and $L_2$ originate from terms of
the type $|A_M\chi|^2$ where $A_M \equiv (A_\mu, \phi_i)$
is the $(4,4)$ vector multiplet in two dimensions. The terms
$L_3$ and $L_4$ arise from commutators of gauge fields in
compactified directions.  

The fields $Y$ are in general
massive. The reason is that the traces $y^{\prime (1)}_{i}
\equiv Y_{i(a'\bar{a}')}^{(1)}$, representing
the centre-of-mass position in the $T^4$ of $Q'_1$
D1-branes, and  $y^{\prime\prime (1)}_{i}
\equiv Y_{i(a''\bar{a}'')}^{(1)}$, representing
the centre-of-mass position in the $T^4$ of $Q''_1$
D1-branes, are neutral and will have {\em vev}'s which
are generically separated (the centres of mass can be
separated in the torus even when they are on top
of each other in physical space). 
 The mass of $Y_{i(a'\bar{a}'')}^{(1)} , 
Y_{i(a''\bar{a}')}^{(1)}$ can be read off from the term $L_3$ in 
\eq{lagrangian}, to be  proportional to
 $ (y^{\prime (1)} - y^{\prime\prime (1)})^2$
Similarly the
mass of $Y_{i(b'\bar{b}'')}^{(5)} , Y_{i(b'\bar{b}'')}^{(5)} $ is
proportional to $(y^{\prime (5)} -y^{\prime\prime (5)})^2 $ 
(as can be read off from the term $L_4$ in \eq{lagrangian})
where $y^{\prime (5)}$ and
$y^{\prime\prime (5) }$ are the centers of mass of the 
$Q'_5$ D5-branes and $Q''_5$
D5-branes along the direction of the dual four torus $\hat{T}^4$. 
(At special points when their centres of mass coincide, these fields
become massless. The analysis for these cases can also be carried out
by incorporating these fields in \eq{dterm}-\eq{resolved}, with no
change in the conclusion) The fields $\phi_{i}^{(1)a'\bar{a}''},
\phi_{i}^{^(1)a''\bar{a}'}$ are also massive. Their masses can be read
off from the $L_1$ in \eq{lagrangian}.
Specifically they arise from the following terms
\be
\chi^*_{a''_1\bar{b}''}\phi_i^{(1)a'\bar{a}''_1*}
\phi_i^{(1)a'\bar{a}''_2}
\chi_{a''_2\bar{b}''}+
\chi^*_{a'_1\bar{b}'}\phi_i^{(1)a''\bar{a}'_1*}\phi_i^{(1)a''\bar{a}'_2}
\chi_{a'_2\bar{b}'} 
\ee ~
where $a'_i$ run from $1,\dots Q'_1$ and
$a''_i$ run form $1,\ldots Q''_1$.  These terms show that
their masses are proportional to the expectation values of the hypers
$\chi_{a'b'}$ and $\chi_{a''b''}$. Similarly the terms of $L_2$ in
\eq{lagrangian} 
\be
\chi^*_{a''\bar{b}''_1}\phi_i^{(5)b''_1{b}'*}\phi_i^{(5)b''_2\bar{b}'}
\chi_{a''\bar{b}''_2}+
\chi^*_{a'\bar{b}'_1}\phi_i^{(5)b'_1\bar{b}''*}\phi_i^{(5)b'_2\bar{b}''}
\chi_{a'\bar{b}'_2} 
\ee 
show that the fields $\phi_{i}^{(5)b'\bar{b}''}
\phi_{i}^{^(5)b''\bar{b}'}$ are massive with masses proportional to the
expectation values of the hypers $\chi_{a'\bar{a}''}$ and
$\chi_{a''\bar{b}''}$. In the above equation $b'_i$ take values
from $1,\ldots , Q'_5$ and $b''_i$ take values from $1,\ldots
, Q''_5$. Note that these masses remain non-zero even in the limit
when the $(Q'_1,Q'_5)$ and $(Q''_1,Q''_5)$ are on the verge of
separating.

Thus the relevant degrees of freedom describing the splitting process
is a $1+1$ dimensional $U(1)$ gauge theory of $A_r$ with $(4,4)$
supersymmetry.  The matter content of this theory consists of
hypermultiplets $\chi_{a'\bar{b}''}$ with charge $+1$ and
$\chi_{a''\bar{b}'}$ with charge $-1$.

Let us now describe the dynamics of the splitting process. This is
given by analyzing the hypermultiplet moduli space of the effective
theory described above with the help of the D-term equations:
\bea
\label{dterm}
A_{a'\bar{b}''}A^*_{a'\bar{b}''}-A_{a''\bar{b'}}A^*_{a''\bar{b}'} - 
B_{b''\bar{a}'}B^*_{b''\bar{a}'}
+B_{b'\bar{a''}}B^*_{b'\bar{a}''} = 0 \\ \nonumber
A_{a'\bar{b}''}B_{b''\bar{a}'} -A_{a''\bar{b}'}B_{b'\bar{a}''} =0
\eea
In the above equations the sum over $a', b', a'' , b''$ is understood.
These equations are generalized version of \eq{dtermr4} 
discussed for the
$R^4/Z_2$ singularity in Section 4.1. 
At the origin of the Higgs branch where the
classical moduli space meets the Coulomb branch this linear sigma model
would flow to an infrared conformal field theory which is singular. 
The reason for this is the same as for the
$R^4/Z_2$ case. The linear sigma model contains the following term in
the superpotential
\be
\label{potential}
V= (A_{a'\bar{b}''}A^*_{a'\bar{b}''} + 
A_{a''\bar{b}'}A^*_{a''\bar{b}'} + B_{b''\bar{a}'}B^*_{b''\bar{a}'} +
B_{b'\bar{a}''}B^*_{b'\bar{a}''})
(\varphi_1^2 + \varphi_2^2 + \varphi_3^2 +\varphi_4^2)
\ee
As in the discussion of the $R^4/Z_2$ case, at the origin of the
hypermultiplet moduli space the flat direction of the Coulomb branch
leads to a ground state which is not normalizable.  This singularity
can be avoided by deforming the D-term equations by
the Fayet-Iliopoulos terms:
\bea
\label{resolved}
A_{a'\bar{b}''}A^*_{a'\bar{b}''}-A_{a''\bar{b}'}A^*_{a''\bar{b}'} -
B_{b''\bar{a}'}B^*_{b''\bar{a}'} +B_{b'\bar{a}''}B^*_{b'\bar{a}''} =
r_3 \\ \nonumber A_{a'\bar{b}''}B_{b''\bar{a}'}
-A_{a''\bar{b}'}B_{b'\bar{a}''} = r_1 + ir_3 
\eea
We note here that
the Fayet-Iliopoulos terms break the relative U(1) under discussion
and the gauge field becomes massive.  The reason is that the D-terms
with the Fayet-Iliopoulos do not permit all $A,B$'s in the above
equation to simultaneously vanish.  At least one of them must be
non-zero, but it is charged, therefore the U(1) is broken. 

The singularity associated with the non-compact Coulomb branch can also
be avoided by turning on the $\theta$ term, the mechanism being
similar to the one discussed in the previous subsection. If any of the
$3$ Fayet-Iliopoulos D-terms or the $\theta$ term is turned on, the
flat directions of the Coulomb branch are lifted, leading to
normalizable ground state is of the Higgs branch. This prevents the
breaking up of the $(Q_1,Q_5)$ system to subsystems. Thus we see that
the $4$ parameters which resolve the singularity of the ${\cal
N}=(4,4)$ SCFT on ${\cal M}$ make their appearance in the gauge theory
as the Fayet-Iliopoulos terms and the theta term.

It would be interesting to extract the singularity structure of the
the gauge theory of the D1/D5 system through mappings similar to
\eq{dtermdef1}- \eq{dtermdef2}.

\subsection{The case $(Q_1,Q_5)\to (Q_1-1,Q_5)+
(1,0)$: splitting of 1 D1-brane}

It is illuminating to consider the special case
in which  1 D1-brane splits off from the bound state
$(Q_1,Q_5)$. The effective dynamics is again 
described in terms of a $U(1)$ gauge theory associated
with the relative separation between the single D1-brane
and the bound state $(Q_1-1,Q_5)$. The massless
hypermultiplets charged under this $U(1)$ correspond
to open strings joining the single D1-brane with
the D5-branes and are denoted by
\be
\chi_{b'} = \left( \begin{array}{c}
A_{b'} \\
B^\dagger_{b'}
\end{array}
\right)
\ee
The D-term equations, with the
Fayet-Iliopoulos terms, become in this case
 \be
\label{dterm1}
\sum_{b'=1}^{Q_5} \left( |A_{b'}|^2 - |B_{b'}|^2 \right)=
r_3, \quad  \sum_{b'=1}^{Q_5} A_{b'} B_{b'} = r_1 + ir_2
\ee
while the potential is
\be
\label{potential1}
V= \left[ \sum_{b'=1}^{Q_5} \left( |A_{b'}|^2 + |B_{b'}|^2 \right)
\right] (\varphi_1^2 + \varphi_2^2 +\varphi_3^2 +\varphi_4^2 )
\ee
The D-term equations above agree with those in
\cite{SeiWit} which discusses the splitting
of a single D1-brane. It is important to emphasize
that the potential and the D-term equations describe
an {\em effective dynamics} in the classical background 
corresponding to the $(Q_1-1, Q_5)$ bound state. This
corresponds to the description in \cite{SeiWit}
of the splitting process in an AdS$_3$ background
which represents a mean field of the above bound state. 

\section{Hawking radiation and the resolutions of the orbifold}

In this section we will address the question whether dynamical
processes like absorption and Hawking radiation from the
five-dimensional black hole are affected by the presence of the above
moduli, especially the blow-up modes.  For our understanding of such
processes to be complete, the supergravity calculation and the SCFT
calculation of absorption cross-section/Hawking radiation rate should
continue to agree even in the presence of these moduli.

\subsection{Supergravity calculation of absorption/Hawking radiation
in presence of moduli}

We recall that the five dimensional black hole solution in the absence
of moduli is \cite{StrVaf,CalMal}\ obtained from the D1-D5 system by
further compactifying $x^5$ on a circle of radius $R_5$ and adding
gravitational waves carrying left-(right-) moving momenta $N_L/R_5$
($N_R/R_5$) along $x^5$. The near horizon geometry of this solution is
\cite{Maldacena,MalStr98}\ BTZ$ \times S^3\times T^4$ where the BTZ
black hole has mass $M=(N_R + N_L)/l$ and angular momentum $J= N_L -
N_R$%
\footnote{$l \equiv 2 \pi (g^2_s Q_1 Q_5/ V_4)^{1/4}$ is the radius of
curvature of the BTZ black hole, $V_4$ being the volume of the
four-torus.}.

The absorption cross-section of minimal scalars in the absence of
moduli is given by
\be
\label{classabs}
\sigma_{abs} = 2 \pi^2 r_1^2 r_5^2 \frac{\pi \om}{2} 
\frac{\exp(\om/T_H)-1}{(\exp(\om/2T_R)-1) (\exp(\om/2T_L)-1)}
\ee

We will now show that the absorption cross-section remains unchanged
even when the moduli are turned on.

{}From the equations of motion of type II supergravity
\cite{CalGubKleTse}, we can explicitly see that the five-dimensional
Einstein metric $ds_{5,Ein}^2$ is not changed by turning on the
sixteen moduli corresponding to the metric $G_{ij}$ on $T^4$ and the
Ramond-Ramond 2-form potential $B'$. As regards the four blowing up
moduli, the invariance of $ds_{5,Ein}^2$ can be seen from the fact
that turning on these moduli corresponds to $SO(4,5)$ transformation
which is part of a U-duality transformation and from the fact that the
Einstein metric does not change under U-duality.  This statement can
be verified by using explicit construction of the corresponding
supergravity solution, at least for the $B_{NS}$ moduli
\cite{DavManWad99}. Now we know that the minimal scalars $\phi^i$ all
satisfy the wave-equation
\be 
D_\mu \del^\mu \phi^i =0 
\ee 
where the Laplacian is with respect to the Einstein metric in five
dimensions. Since it is {\em only} this  wave equation that
determines the absorption cross-section completely, we see that
$\sigma_{abs}$ is the same as before.

It is straightforward to see that the Hawking rate, given by
\be
\label{classdecay}
\Gamma_H = \sigma_{abs} (e^{\om/T_H}-1)^{-1} 
\frac{d^4k}{(2 \pi)^4}
\ee
is also not changed.

\subsection{SCFT calculation of absorption cross-section/Hawking
rate in the presence of moduli}

In Section 3 we have listed the twenty (1,1) operators $O_i(z,\bar z)$
in the SCFT based on the symmetric product orbifold ${\cal M}$ which
is dual to the D1/D5 system.  In order to arrive at the SCFT dual to
the {\em black hole}, we have to first implement the periodic
identification $x^5 \equiv x^5 + 2\pi R_5$.  It was shown in
\cite{MalStr98} that this forces the SCFT to be in the Ramond sector.
Turning on  various moduli $\phi^i$ of supergravity
corresponds to perturbing the SCFT
\be
\label{CFTperturbation}
S= S_0 +  \sum_i \int d^2 z \bar \phi^i O_i(z,\bar z) 
\ee
where $\bar \phi^i$ denote the near-horizon limits of the
various moduli fields $\phi^i$.

Finally, the Kaluza-Klein momentum of the black hole (equivalently,
angular momentum of BTZ) corresponds to excited states of this sector
with 
\be
\label{KKmom}
L_0=N_L, \bar L_0 = N_R
\ee  

We note here that $S_0$ corresponds to the free SCFT based on the
symmetric product orbifold ${\cal M}$. Since this SCFT is non-singular
(all correlation functions are finite), it does not correspond to the
marginally stable BPS solution originally found in
\cite{StrVaf,CalMal}.  Instead, it corresponds to a five-dimensional
black hole solution in supergravity with suitable ``blow-up'' moduli
turned on.

Let us now calculate the absorption cross-section of a supergravity
fluctuation $\delta \phi_i$ using SCFT.  The notation $\delta\phi_i$
implies that we are considering the supergravity field to be of the
form
\be
\phi^i = \phi^i_0 + \kappa_5\delta \phi^i
\ee 
where $\phi^i_0$ represents the background value. The factor of
$\kappa_5$ above ensures appropriate normalization of the
fluctuation $\delta \phi^i$ as explained below in
the paragraph above \eq{full_action}. 
This corresponds to the SCFT action
\bea
\label{CFTperturbation1}
S &=& S_0 + \int d^2 z [\bar \phi^i_0 + \kappa_5 \delta \bar 
\phi^i] O_i(z,\bar z)
\nonumber \\
&=& S_{\phi_0} + S_{int}
\nonumber \\
\eea
where
\be
\label{S_phi_0}
S_{\phi_0} = S_0 + \int d^2 z \bar \phi_0^i O_i(z, \bar z)
\ee
\be
\label{S_int}
S_{int} = \kappa_5 \int d^2 z \delta \bar \phi^i O_i(z, \bar z)
\ee

The absorption cross-section of the supergravity fluctuation
$\delta\phi^i$ involves \cite{Gubser,DhaManWad,DavManWad} essentially
the two-point function of the operator $O_i$ calculated with respect
to the SCFT action $S_{\phi_0}$. Since $O_i$ is a marginal operator,
its two-point function is determined apart from a
normalization constant.

Regarding the marginality of the operators $O_i$, it is easy to
establish it upto one-loop order by direct computation
($c_{ijk}=0$). The fact that these operators are exactly marginal can
be argued as follows. The twenty operators $O_i$ arise as top
components of five chiral primaries. It is known that the number of
chiral primaries with $(j_R, \tilde j_R)=(m,n)$ is the Hodge number
$h_{2m,2n}$ of the target space ${\cal M}$ of the SCFT. Since this
number is a topological invariant, it should be the same at all points
of the moduli space of deformations.

In the case studied in \cite{DavManWad}\ it was found that if the
operator $O_i$ is canonically normalized (OPE has residue 1) and if
$\delta \phi_i$ is canonically normalized in supergravity, then the
normalization of $S_{int}$ as in \eq{S_int} ensures that
$\sigma_{abs}$ from SCFT agrees with the supergravity result.  The
crucial point now is the following: once we fix the normalization of
$S_{int}$ at a given point in moduli space, at some other point it may
acquire a constant ($\not=1$) in front of the integral when $O_i$ and
$\delta\phi_i$ are canonically normalized at the new point. This would
imply that $\sigma_{abs}$ will get multiplied by this constant, in
turn implying disagreement with supergravity.  We need to show that
that does not happen.
 
To start with a simple example, let us first restrict to the moduli
$g_{ij}$ of the torus $\widetilde{T^4}$. We have
\be
\label{full_action}
S = \int d^2 z \del x^i \bar \del x^j g_{ij}
\ee
The factor of string tension has been absorbed in
the definition of $x^i$.

In \cite{DavManWad} we had $g_{ij}
= \delta_{ij} +  \kappa_5 h_{ij}$, leading to
\bea
S &=& S_0 + S_{int} \nonumber\\
S_0 &=&  \int d^2 z \del x^i \bar \del x^j \delta_{ij}
\nonumber\\
S_{int} &=& \kappa_5  
\int d^2 z \del x^i \bar \del x^j h_{ij}
\nonumber \\
\eea
As we remarked above, this $S_{int}$ gives rise to the
correctly normalized $\sigma_{abs}$. 

Now, if we expand around some other metric 
\be
g_{ij} = g_{0ij} + \kappa_5 h_{ij}
\ee
the above action \eq{full_action} implies 
\bea
S &=& S_{g_0} + S_{int} \nonumber\\
S_{g_0} &=&  \int d^2 z \del x^i \bar \del x^j g_{0ij}
\nonumber\\
S_{int} &=& \kappa_5 
\int d^2 z \del x^i \bar \del x^j h_{ij}
\nonumber \\
\eea
Now the point is, neither $h_{ij}$ nor the operator $O^{ij} =\del X^i
\bar \del X^j $ in $S_{int}$ is canonically normalized at $g_{ij}=
g_{0,ij}$. When we do use the canonically normalized operators, do we
pick up an additional constant in front?

Note that 
\be
\label{zamol_example}
\langle O^{ij} O^{kl} \rangle_{g_0}=g_0^{ik} g_0^{jl}|z-w|^{-4}
\ee
and
\be
\label{sugra_propagator_example}
\langle h_{ij} (x) h_{kl} (y) \rangle_{g_0} =
g_{0,ik} g_{0,jl} {\cal D}(x,y)
\ee
${\cal D}(x,y)$ is the massless scalar propagator. 

This shows that

Statement (1):{\em The two-point functions of $O^{ij}$ and $h_{ij}$
pick up inverse factors.}

As a result, $S_{int}$ remains correctly normalized when re-written
in terms of the canonically normalized $h$ and $O$, and no additional
constant is picked up.

The above result is in fact valid in the full twenty dimensional
moduli space $\tM$ because Statement (1) above remains true
generally. 

To see this, let us first rephrase our result for the special case of
the metric moduli \eq{full_action} in a more geometric way.  The
$g_{ij}$'s can be regarded as some of the coordinates of the moduli
space $\tM$ (known to be a coset $SO(4,5)/(SO(4)\times SO(5))$).  The
infinitesimal perturbations $h_{ij}, h_{kl}$ can be thought of as
defining tangent vectors at the point $g_{0,ij}$ (namely the vectors
$\del/\del g_{ij}, \del/\del g_{kl}$). The (residue of the) two-point
function given by \eq{zamol_example}\ defines the inner product
between these two tangent vectors according to the Zamolodchikov
metric \cite{Zamolodchikov}.

Consider, on the other hand, the propagator (inverse two-point
function) of $h_{ij}, h_{kl}$ in supergravity. The moduli space action
of low energy fluctuations is nothing but the supergravity action
evaluated around the classical solutions $g_{0,ij}$.  The kinetic term
of such a moduli space action defines the metric of moduli space. The
italicized statement above is a simple reflection of the fact that the
Zamolodchikov metric defines the metric on moduli space, and hence 

Statement (2): {\em the propagator of supergravity fluctuations, viewed
as a matrix, is the inverse of the two-point functions in the SCFT.}

The last statement is of course not specific to the moduli $g_{ij}$
and is true of all the moduli.  We find, therefore, that fixing the
normalization of $S_{int}$ \eq{S_int} at any one point $\phi_0$
ensures that the normalization remains correct at any other point
$\phi'_0$ by virtue of Statement (2). We should note in passing that
Statement (2) is consistent with, and could have been derived from,
AdS/CFT correspondence as applied to the two-point function.

Thus, we find that $\sigma_{abs}$ is independent of the
moduli, in agreement with the result from supergravity.

\subsection{Entropy and area}

We make a brief mention here of the fact that the correspondence
between Bekenstein-Hawking entropy and the SCFT entropy remains true
in the presence of all the twenty moduli. The reason is that in
supergravity the Einstein metric remains unchanged (see Section 5.1)
and therefore the area of the event horizon remains the same (this can
be explicitly verified using the supergravity solution in
\cite{DavManWad99}).  In the SCFT, since the operators corresponding to
the above moduli are all exactly marginal (Section 5.2) therefore the
central charge remains unchanged and hence, by Cardy's formula, the
entropy does not change, in agreement with the Bekenstein-Hawking
formula.

\section{Conclusions}

(a) We presented an explicit construction of all the marginal
operators in the SCFT of the D1/D5 system based on the orbifold ${\cal
M}$. These are twenty in number, four of which are constructed using
$Z_2$ twist operators and correspond to blowing up modes of the
orbifold.

\noindent
(b) We classified the the twenty near-horizon moduli of supergravity
on AdS$_3 \times S^3 \times T^4$ according to representations of
$SU(1,1|2) \times SU(1,1|2) \times SO(4)_I$.

\noindent
(c) We established one-to-one correspondence between the supergravity
moduli and the marginal operators by inventing a new $SO(4)$ symmetry
in the SCFT which we identified with the $SO(4)_I$ of supergravity.

\noindent
(d) We analyzed gauge theory dynamics of the D1/D5 system relevant for
the splitting of the bound state $(Q_1,Q_5) \to (Q'_1,Q'_5) +
(Q''_1,Q''_5)$.

\noindent
(e) We showed in supergravity as well as in SCFT that the absorption
cross-section for minimal scalars is the same for all values of the
moduli, therefore establishing the agreement between SCFT and
supergravity all over the moduli space.

\vspace{15ex}

{\bf Acknowledgement}: We would like to thank Avinash Dhar and
K.S. Narain for discussions.  S.W. would like to thank the ASICTP for
their warm hospitality.

\appendix

\section{The ${\cal N} =4$ superconformal algebra}

To set up our notations and conventions we review the ${\cal N}=4$
superconformal algebra. The algebra is generated by the stress energy
tensor, four supersymmetry currents, and a local $SU(2)$ $R$ symmetry
current. The operator product expansions of the algebra 
with central charge $c$ are given by (See for example \cite{Yu}.)
\bea
T(z)T(w) &=& \frac{\del T(w)}{z-w} + \frac{2 T(w)}{(z-w)^2} +
\frac{c}{2 (z-w)^4},  \\  \nonumber
G^a(z)G^{b\dagger }(w) &=& 
\frac{2 T(w)\delta_{ab}}{z-w} + \frac{2 \bar{\sigma}^i_{ab} \del J^i}
{z-w} + \frac{ 4 \bar{\sigma}^i_{ab} J^i}{(z-w)^2} + 
\frac{2c\delta_{ab}}{3(z-w)^3}, \\
\nonumber
J^i(z) J^j(w) &=& \frac{i\epsilon^{ijk} J^k}{z-w} + \frac{c}{12 (z-w)^2}
, \\  \nonumber
T(z)G^a(w) &=& \frac{\del G^a (w)}{z-w} + \frac{3 G^a (z)}{2 (z-w)^2},
\\   \nonumber
T(z) G^{a\dagger }(w) &=& \frac{\del G^{a\dagger} (w)}{z-w} + 
\frac{3 G^{a\dagger} (z)}{2 (z-w)^2}, \\   \nonumber
T(z) J^i(w) &=& \frac{\del J^i (w)}{z-w} + \frac{J^i}{(z-w)^2}, \\
\nonumber
J^i(z) G^a (w) &=& \frac{G^b(z) (\sigma^i)^{ba}}{2 (z-w)}, \\
\nonumber
J^i(z) G^{a\dagger}(w) &=& -\frac{(\sigma^i)^{ab} G^{b\dagger
}(w)}{2(z-w)} 
\eea
Here $T(z)$ is the stress energy tensor, $G^a(z), G^{b\dagger}(z)$ 
the  $SU(2)$
doublet of supersymmetry generators and  $J^i(z)$ the $SU(2)$ $R$
symmetry current. The $\sigma$'s stand for Pauli matrices.
In the free field realization desribed below,
the above holomorphic currents occur together with
their antiholomorphic counterparts, which we will
denote by $\widetilde J(\bar z), \widetilde G(\bar z)$
and $\widetilde T(\bar z)$. In
particular, the R-parity group will be denoted
by $SU(2)_R \times \widetilde{SU(2)}_R$.

A free field realization of the ${\cal N}=4$ superconformal 
algebra with $c=6Q_1Q_5$
can be constructed out of $Q_1Q_5$ copies 
of four real fermions and bosons. 
The generators are given by
\bea
T(z) &=& \del X_A (z)\del X^\dagger_A(z) 
+ \frac{1}{2}\Psi_A (z)\del \Psi^{\dagger}_A (z) 
- \frac{1}{2}\del\Psi_A (z) \Psi^{\dagger}_A (z) 
\\   \nonumber
G^a(z) &=&
\left(
\begin{array}{c}
G^1(z)\\
G^2(z)
\end{array}
\right) =
\sqrt{2} \left(
\begin{array}{c}
\Psi^1_A (z) \\
\Psi^2_A (z) \end{array}  \right)  \del X^2_A (z)  +
\sqrt{2} \left( \begin{array}{c}
-\Psi^{2\dagger}_A (z) \\
\Psi^{1\dagger}_A (z)  \end{array}  \right)  \del X^1_A (z) 
\\    \nonumber
J^i_R(z) &=& \frac{1}{2} \Psi_A(z)\sigma^i\Psi^\dagger_A (z)\\
\nonumber
\eea
We will use the following notation for the zero mode
of the R-parity current:
\be
J^i_R = \frac{1}{2}\int\frac{dz}{2\pi i} 
\Psi_A(z)\sigma^i\Psi^\dagger_A(z)
\ee
In the above the summation over $A$ which runs from $1$ to $Q_1Q_5$ is
implied.
The bosons $X$ and the fermions $\Psi$ are 
\bea
\label{defn}
X_A(z) &=& (X^1_A(z), X^2_A(z)) = \sqrt{1/2} (x^1_A(z) + i x^2_A(z),
x^3_A(z) + i x^4_A(z)),   \\  \nonumber
\Psi_A (z) &=& (\Psi^1_A(z), \Psi^2_A(z)) = \sqrt{1/2} (\psi^1_A(z) +
i\psi^2_A(z), \psi^3_A(z) + i\psi^4_A(z)) \\   \nonumber
X_A^\dagger (z) &=& 
\left(
\begin{array}{c}
X_A^{1\dagger} (z) \\
X_A^{2\dagger} (z)
\end{array}
\right)   = \sqrt{\frac{1}{2}}
\left(
\begin{array}{c}
x^1_A(z)-ix^2_A (z)\\
x^2_A(z)-ix^2_A(z)
\end{array}
\right) \\   \nonumber
\Psi_A^\dagger (z) &=&
\left(
\begin{array}{c}
\Psi_A^{1\dagger} (z) \\
\Psi^{2_A \dagger} (z)
\end{array}
\right)     =\sqrt{\frac{1}{2}} 
\left(  
\begin{array}{c}
\psi^1_A(z) - i\psi^2_A(z)\\
\psi^3_A(z) - i\psi^4_A(z)
\end{array}
\right)
\eea

In addition to the local $R$ symmetry the free field realization of
the ${\cal N}=4$ superconformal algebra has additional global
symmetries which can be used to classify the states. There are $2$
global $SU(2)$ symmetries which correspond to the $SO(4)$ rotations of
the $4$ bosons $x^i$. The corresponding charges are given by 
\bea
I_1^i &=& 
\frac{1}{4}\int\frac{dz}{2\pi i} X_A \sigma^i \del X_A^\dagger 
-\frac{1}{4}\int\frac{dz}{2\pi i} \del X_A \sigma^i X_A^\dagger 
+ \frac{1}{2}\int\frac{dz}{2\pi i}
\Phi_A \sigma^i \Phi_A^\dagger  \\  \nonumber
I_2^i &=& 
\frac{1}{4} 
\int\frac{dz}{2\pi i}{\cal X}_A\sigma^i\del{\cal X}_A^\dagger
-\frac{1}{4} 
\int\frac{dz}{2\pi i}\del{\cal X}_A\sigma^i{\cal X}_A^\dagger
\eea
Here 
\bea
{\cal X }_A = (X^1_A, -X^{2\dagger}_A) \;&\;&\;\;
{\cal X}^\dagger  =
\left(
\begin{array}{c}
X^{1\dagger}_A \\
-X^2_A 
\end{array}  \right) \nonumber \\
\Phi_A = (\Psi^1_A, \Psi^{2\dagger}_A ) \;&\;&\;\;
\Phi_A^\dagger =
\left(
\begin{array}{c}
\Psi^{1\dagger}_A  \\
\Psi^2_A 
\end{array} \right).
\eea
These charges are generators of $SU(2)\times SU(2)$ algebra:
\bea
[I_1^i, I_1^j] = i\epsilon^{ijk} I_1^k \;&\;&\;\;
[I_2^i, I_2^j] = i\epsilon^{ijk} I_2^k 
\\  \nonumber
[I_1^i, J_2^j] &=&0
\eea
The commutation relation of these new global charges with the various
local charges are given below
\bea
\label{so4_on_g}
[I_1^i, G^a(z)] =0 \;&\;&\;\;
[I_1^i, G^{a\dagger}(z)] =0 \\ \nonumber 
[I_1^i, T(z)] =0 \;&\;&\;\;
[I_1^i, J(z)]=0 \\ \nonumber 
[I_2^i, {\cal G}^a(z)] = 
\frac{1}{2}{\cal G}^{b} (z)\sigma^i_{ba} \;&\;&\;\;
[I_2^i, {\cal G}^{a\dagger} (z) ]
= - \frac{1}{2}\sigma^i_{ab}{\cal G}^{b\dagger}(z) \\ \nonumber
[I_2^i, T(z)] =0  \;&\;&\;\;
[I_2^i, J(z)]=0
\eea
where
\bea
{\cal G} = ( G^1, G^{2\dagger}) \;&\;&\;\;
{\cal G}^\dagger=
\left(
\begin{array}{c}
G^{1\dagger} \\
G^2
\end{array}
\right)
\eea

The following commutations relation show that 
the bosons transform as $(\bf 2, \bf 2)$ under $SU(2)_{I_1}\times
SU(2)_{I_2}$
\bea
\label{boson}
[I_1^i, X^a_A] = \frac{1}{2} X^b_A\sigma^i_{ba} \;&\;&\;\;
[I_1^i, X^{a\dagger}_A] = -\frac{1}{2}\sigma^i_{ab}X^{b\dagger}_A 
\\   \nonumber
[I_2^i, {\cal X}^a_A ] =
\frac{1}{2} {\cal X}^b_A \sigma^i_{ba}  \;&\;&\;\;
[I_2^i, {\cal X}^{a\dagger}_A] =
-\frac{1}{2}\sigma^i_{ab}{\cal X}^{b\dagger}_A 
\eea
The fermions transform as $(\bf 2, \bf 1)$ under $SU(2)_{I_1}\times 
SU(2)_{I_2}$ as can be seen from the commutations relations
given below.
\bea
[I_1^i, \Phi^a_A] = \frac{1}{2}\Phi^b_A \sigma^i_{ba} \;&\;&\;\;
[I_1^i, \Phi^{a\dagger}_A] =
-\frac{1}{2}\sigma^i_{ab}\Phi^{b\dagger}_A  \\   \nonumber
[I_2^i, \Psi^a] =0  \;&\;&\;\;
[I_2^i, \bar{\Psi}^a]=0
\eea
We are interested in  studying the states of the ${\cal N}=(4,4)$ SCFT
on ${\cal M}$. The classification of the states and their symmetry
properties can be analyzed by studying the states of a free field
realization of a ${\cal N}=(4,4)$ SCFT on $R^{4Q_1Q_5}/S(Q_1Q_5)$. 
This
is realized by considering the holomorphic and the anti-holomorphic
${\cal N}=4$ superconformal algebra with $c=\bar{c}=6Q_1Q_5$
constructed out of $Q_1Q_5$ copies of four real fermions and bosons. 
So we have an anti-holomorphic component for each field, generator and
charges discussed above. These are labelled by the same symbols used
for the holomorphic components but distinguished by a tilde.

The charges $I_1, I_2$ constructed
above  generate $SO(4)$
transformations only on  the {\em holomorphic} bosons $X_A(z)$. 
Similarly, we can construct charges
$\widetilde{I_1}, \widetilde{I_2}$ which generate $SO(4)$
transformations only on  the {\em antiholomorphic} 
bosons $\widetilde{X_A}(\bar z)$. 
Normally one would expect these
charges to give rise to a global $SO(4)_{hol}
\times SO(4)_{antihol}$ symmetry. However, 
the kinetic term of the bosons in the
free field realization is not invariant under independent holomorphic
and antiholomorphic $SO(4)$ rotations. It is
easy to see, for example by using the Noether
procedure, that there is a residual $SO(4)$ symmetry
generated by the charges 
\bea
J_I= I_1  + \widetilde{I}_1 \;\;&\;&\;
\widetilde{J}_I = I_2 + \widetilde{I}_2
\eea
We will denote this symmetry as $SO(4)_I =
SU(2)_I\times \widetilde{SU(2)}_I$, where the
$SU(2)$ factors are generated by $J_I,
\widetilde{J}_I$. These charges satisfy the
property that (a) they 
correspond to $SO(4)$ transformations of the
bosons $X_A(z,
\bar z)= X_A(z) 
+ \widetilde{X_A}(\bar z)$ and (b) they fall into representations of the
${\cal{N}}=(4,4)$ algebra (as can 
be proved by using the commutation relations \eq{boson}
of the $I$'s). The bosons $X(z,\bar{z})$
transform as $(\bf 2 , \bf 2)$ under $SU(2)_I\times \widetilde{SU(2)}_I$. 

\section{Short multiplets of $SU(1,1|2)$}

The supergroup $SU(1,1|2)$ is the global part of the ${\cal N}=4$
superconformal algebra. 
The representations of this
supergroup are classified according to the conformal weight 
and $SU(2)_R$ quantum number. The highest weight states
$ |\mbox{hw}\rangle = 
|h,{\bf j}_R,j_R^3 =j_R \rangle $ satisfy the following
properties
\bea
L_1 |\mbox{hw}\rangle = 0 \;\;\; 
L_0 |\mbox{hw}\rangle = h|\mbox{hw}\rangle \\ \nonumber
J^{(+)}_{R}|\mbox{hw}\rangle =0  \;\;\; 
J_R^{(3)}|\mbox{hw}\rangle = j_R|\mbox{hw}\rangle\\  \nonumber
G_{1/2}^a|\mbox{hw}\rangle =0 \;\;\; G_{1/2}^{a\dagger}
|\mbox{hw}\rangle =0
\eea
Where $L_{\pm,0} ,J^{(\pm),(3)}_R$ are the global charges of the currents
$T(z)$ and $J^{(i)}_R(z)$. The charges $G^a_{1/2,-1/2} $ are 
the global charges of the supersymmetry currents $G^a(z)$ 
in the Neveu-Schwarz sector. Highest
weight states which satisfy 
$
G^{2\dagger }_{-1/2}|\mbox{hw}\rangle =0 ,\;\;\;
G^1_{-1/2}|\mbox{hw}\rangle =0
$
are chiral primaries. They satisfy $h=j$. We will denote 
these states as $|\mbox{hw}\rangle _{S}$. Short multiplets are
generated from the chiral primaries through the action of the raising
operators $J_{-}, G^{1\dagger }_{-1/2}$ and $G^2_{-1/2}$. The structure
of the short multiplet is given below
\bea
\label{short}
\begin{array}{cccc}
\mbox{States} & j & L_0 & \mbox{Degeneracy} \\
|\mbox{hw}\rangle_{S} & h & h& 2h+1 \\
G^{1\dagger }_{-1/2}|\mbox{hw}\rangle_{S}, 
G^2_{-1/2}|\mbox{hw}\rangle_{S} &h-1/2& h+1/2 & 2h + 2h = 4h \\
G^{1\dagger }_{-1/2} G^2_{-1/2}|\mbox{hw}\rangle_{S} & h-1& h+1& 2h-1
\end{array}
\eea
The short multiplets of the supergroup $SU(1,1|2)\times SU(1,1|2)$ are
obtained by the tensor product of the above multiplet. We denote the
short multiplet of  $SU(1,1|2)\times SU(1,1|2)$ as
$(\bf{2h +1}, \bf{2h'+1})_S$. These stand for the degeneracy of the
bottom component, the top row in  \eq{short}. The top component of
the short multiplet are the states belonging to  the last row in
\eq{short}. The short multiplet $(\bf{2}, \bf{2})_S$ is special, it
terminates at the middle row of \eq{short}. For this case, the top
component is the middle row. These states have $h=\bar{h}=1$
and transform as $(\bf{1}, \bf{1})$ of $SU(2)_R\times
\widetilde{SU(2)}_R$. There are $4$ such states for each $(\bf{2},
\bf{2})_S$.

\end{document}